\documentclass[%
 preprint,
 longbibliography,
 amsmath,amssymb,
 aps,
 pra,
]{revtex4-2}
\usepackage[T1]{fontenc} 
\usepackage[utf8]{inputenc}
\usepackage{graphicx}
\usepackage{dcolumn}
\usepackage{siunitx}
\usepackage{bm}
\usepackage{hyperref}
\usepackage{amssymb}
\usepackage{mathtools}
\usepackage{amsmath}
\usepackage{makecell}
\usepackage{float}
\makeatletter
\let\newfloat\newfloat@ltx
\makeatother
\usepackage{algorithm}
\usepackage{algpseudocode}
\usepackage{tabularx}
\usepackage{tikz}
\usepackage[nolist]{acronym}
\usepackage[nameinlink,capitalise]{cleveref}
\crefname{algorithm}{Alg.}{Algs.}
\Crefname{algorithm}{Algorithm}{Algorithms}
\crefname{table}{Tab.}{Tabs.}
\Crefname{table}{Table}{Tables}
\crefname{section}{Sec.}{Secs.}
\Crefname{table}{Section}{Sections}

\begin{document}
\begin{acronym}[DIKW-hierarchy]\itemsep0pt
\acro{MNIST}{Modified National Institute of Standards and Technology}
\acro{CNN}{Convolutional Neural Network}
\acro{GAN}{Generative Adversarial Network}
\acro{FPGA}{Field Programmable Gate Array}
\acro{GPU}{Graphics Processing Unit}
\acro{LINAC}{Linear Accelerator}
\acro{FEL}{Free Electron Laser}
\acro{BESSY II}{Berliner Elektronenspeicherring für Synchrotronstrahlung II}
\acro{EMIL}{Energy Materials In situ Laboratory}
\acro{ERL}{Energy Recovery \acf{LINAC}}
\acro{VAE}{Variational Autoencoder}
\acro{SMA}{Shape Memory Alloy}
\acro{ODF}{Orientation Density Functions}
\acro{GPU}{Graphics Processing Unit}
\acro{WIMV}{Williams, Imhof, Matthies and Vinel}
\acro{MAE}{Mean Absolute Error}
\acro{MSE}{Mean Squared Error}
\acro{RMSE}{Rooted Mean Squared Error}
\acro{ELU}{Exponential Linear Unit}
\acro{MLP}{Multilayer Perceptron}
\acro{CAE}{Convolutional Autoencoder}
\acro{CPU}{Central Processing Unit}
\acro{CI}{Confidence Interval}
\acro{CCNN}{Circular Convolutional Neural Network}
\acro{XFEL}{X-ray Free Electron Laser}
\acro{LCLS}{\acf{LINAC} Coherent Light Source}
\acro{TOF}{time-of-flight}
\acro{SRF}{Superconducting Radiofrequency}
\acro{RL}{Reinforcement Learning}
\acro{ASTRA}{A Space Charge Tracking Algorithm}
\end{acronym}

\title{Reconstructing Time-of-Flight Detector Values of Angular Streaking Using Machine Learning}

\author{David~Meier}
\affiliation{%
Helmholtz-Zentrum f\"ur Materialien und Energie GmbH, Albert-Einstein-Straße 15, 12489 Berlin, Germany
}%

\author{Wolfram Helml}%
\affiliation{%
Zentrum für Synchrotronstrahlung, Technische Universität Dortmund, 44227 Dortmund, Germany
}

\author{Thorsten Otto}%
\affiliation{%
Deutsches Elektronen-Synchrotron DESY, Notkestr. 85, 22607 Hamburg, Germany
}

\author{Bernhard~Sick}%
\affiliation{%
Intelligent Embedded Systems, University of Kassel, Wilhelmsh\"oher Allee 73, 34121 Kassel, Germany
}%
\author{Jens~Viefhaus}%
\author{Gregor~Hartmann}%
\affiliation{%
Helmholtz-Zentrum f\"ur Materialien und Energie GmbH, Albert-Einstein-Straße 15, 12489 Berlin, Germany
}%

\collaboration{AIM-ED --
Joint Lab Helmholtz-Zentrum f\"ur Materialien und Energie, Berlin (HZB) and
University of Kassel}\noaffiliation

\date{\today}

\begin{abstract}
Angular streaking experiments enable for experimentation in the attosecond regions. However, the deployed \ac{TOF} detectors are susceptible to noise and failure. These shortcomings make the outputs of the \ac{TOF} detectors hard to understand for humans and further processing, such as for example the extraction of beam properties. In this 
article, we present an approach to remove high noise levels and reconstruct up to three failed \ac{TOF} detectors from an arrangement of 16 \ac{TOF} detectors. Due to its fast evaluation time, the presented method is applicable online during a running experiment. It is trained with simulation data, and we show the results of denoising and reconstruction of our method on real-world experiment data.
\end{abstract}

\maketitle


\section{Introduction}
\label{sec:introduction_tof}
During the last years, the \acp{XFEL} have been developed and established as a new generation of light source for X-ray experiments \cite{Decking2020}. The \ac{LCLS} located in Stanford, California, is such a light source \cite{Bostedt2016}. \acp{XFEL} can produce radiation with nearly ten orders of magnitude higher brightness compared to previous light sources. The pulse durations can range from 500 to less than subfemtoseconds ($10^{-15} s$) \cite{Hartmann2018}. This property is beneficial because most experiments performed at \acp{XFEL} require a short, well-defined photon pulse.

At \ac{LCLS}, angular streaking experiments are carried out. The main goal of these experiments are measurements in the attosecond ($10^{-18} s$) region. These experiments allow us to gain previously inaccessible knowledge, since some atomic processes happen on this time scale. For example, tunnel ionization is an effect where a particle goes through a potential barrier. A measuring resolution in attoseconds is required to analyze this phenomenon and has successfully been conducted \cite{Sainadh2019}. In angular streaking, X-rays ionize the gas introduced into the chamber through a gas needle for examination. That means that so-called photoelectrons are emitted from this gas. A circularly polarized laser
has temporal and spatial overlap with the X-ray pulses in the target gas chamber. The laser's electromagnetic field then streaks the emitted photoelectrons according to their emission time in different angular positions due to the circular polarization of the laser. In our particular experimental setup, the emitted electrons are detected by an angularly aligned array of 16 electron \ac{TOF} spectrometers set up in the dipole plane, i.e. the plane perpendicular to the incoming X-ray pulses. The \ac{TOF} spectrometers are located in steps of \ang{22.5} and are thus covering the entire circle of the dipole plane. For a detailed description on the experimental setup and how the real experiment data have been recorded, please refer to \cite{Hartmann2018}.

The \ac{TOF} spectrometers used in the experiment are highly sensitive, and an unknown number of \ac{TOF} detectors can break before or during the experiment. They can fail completely or produce implausible results. If they produce implausible results, they are disabled manually. While automating the detection of such events using anomaly detection methods could be a valuable extension, it is not covered within the scope of this study. In addition to frequent failures, the \ac{TOF} detectors suffer from extremely high noise. Additionally, to avoid saturation, these detectors are restricted in the number of electrons they collect. However, this constraint results in an insufficient sample size for accurate Poisson statistics.

Therefore, we present an approach to reduce noise and reconstruct the information of failed \ac{TOF} detectors using simulation knowledge and neural networks and demonstrate this approach with real-world data from the experiment at \ac{LCLS}. For this approach, we train a neural network exclusively on simulated data. This neural network receives the simulated detector images augmented with artificially added noise and deactivated \ac{TOF} detectors mimicking failed \ac{TOF} detectors. The neural network is trained to map these noisy, incomplete images to their corresponding noise-free versions with no failed \ac{TOF} detectors. Since the training data is fully simulated, the ground truth for non-noisy and complete detector images is known, enabling evaluation and refinement of the network's outputs during the training process.

The main advances provided by the method proposed in the present study are:
\subsection{Fast Denoising of \texorpdfstring{\ac{TOF}}{TOF} Detector Data}
We propose a neural network-based approach for high-noise reduction in detector images, enhancing readability and preparing the data for further automated analysis. This approach is specifically designed for \textit{online} processing, meaning that it adds a maximum delay of one second per data batch. It handles data with batch sizes of at least 1024 detector images in $217.53 ~\mathrm{ms}$ with $100$ threads on an AMD EPYC 7662, allowing it to match with one second data batches at \ac{LCLS} experiment, which supplies data at a rate of 120 detector images per second. This capability enables near real-time feedback for the experimentator and further online processing, similar to the retrospective evaluation conducted in \cite{Hartmann2018}. Additionally, extracting characteristics online from the X-ray pulse is possible, as presented in \cite{Dingel2022} for a similar experiment at European \ac{XFEL} located in Hamburg, Germany. 

\subsection{Fast Reconstruction of Missing \texorpdfstring{\ac{TOF}}{TOF} Detectors Using Simulation Knowledge}
Furthermore, we want to reconstruct missing \ac{TOF} detectors. We assume that defective \ac{TOF} detectors either fail completely or have all their channels manually set to zero in the event of incorrect outputs. We will train neural networks that can reconstruct up to three failed \ac{TOF} detectors. We compare networks trained explicitly for a distinct number of failed \ac{TOF} detectors with those trained for one up to three failed \ac{TOF} detectors. This reconstruction capability is also crucial for further processing steps, with our method allowing reconstructions online, as previously defined.

\subsection{Improving Future Designs of \texorpdfstring{\ac{TOF}}{TOF} Arrays}
Because our neural network models are trained only using simulation data, we know which information was left out for reconstruction. Based on the reconstruction quality of our \ac{TOF} reconstruction approach, we can determine at which points our reconstruction algorithm has difficulties. By that, these \ac{TOF} arrays presumably have less redundant information. This knowledge can be used to improve future designs of \ac{TOF} arrays, for example, by having a more significant overlap of specific \ac{TOF} detectors or by constructing \ac{TOF} arrays entirely different. For instance, it could be beneficial to use a different geometry, to avoid the loss of too much redundant information if two correlated \ac{TOF} detectors fail.

\section{Related Work}
\label{sec:tof_rel_work}
Since the data recorded from the \ac{TOF} detectors are two-dimensional, we consider them as images in this discussion. In the literature, the task of reconstructing missing parts of images is referred to as \textit{image inpainting}. Typically, parts of the image are distorted or covered, and the distorted or concealed parts should be recovered. One existing method is the simple and fast \textit{coherence transport} \cite{Bornemann2007}. These algorithms are based on nonlinear higher-order partial differential equations. Nevertheless, these methods are only based on a single image and cannot incorporate knowledge from a simulation or image database. Furthermore, they rely on containing enough information on this single image to recover the missing parts of the image.

More sophisticated approaches incorporating knowledge from image databases containing images from the same domain use \acp{GAN}. The following summary on the development of \ac{GAN} approaches is based on \cite{Hukkelas2021}. A significant advance was using \acp{CNN} for image inpainting \cite{Koehler2014, Xie2012}. However, these methods could not distinguish if an input pixel exists in the data or is missing. The approach of using \textit{partial convolution} is a modification that replaces missing input pixels with zeros and normalizes the output depending on the number of valid pixels \cite{Liu2018}. This approach, however, relies on manual hardcoded certainty propagation. There exist methods to replace these components by learning them. Nevertheless, they use about half of the network parameters to propagate the certainties through the network \cite{Xie2019, Yu2019}. This makes the evaluation of the networks more expensive. Furthermore, \ac{GAN}-based approaches are hard to train because they typically suffer from mode collapse \cite{Kossale2022}. Mode collapse means the generator repeatedly produces the same image since it is a plausible output for the discriminator. If the discriminator fails to learn to reject this sample because it is in a local minimum, the generator-discriminator loop is stuck to this image. Moreover, \acp{GAN} can suffer from vanishing gradients when the discriminator is trained too well \cite{Arjovsky2017}. Some of these shortcomings can be mitigated using the so-called Wasserstein-\acp{GAN}, which uses the Wasserstein metric to measure the distance of the latent variable distribution from a standard normal distribution \cite{Arjovsky2017}. However, they are slower in the training process because the Wasserstein distance calculation is computationally more expensive.

More recently, so-called stable diffusion models, initially designed for text-to-image transformations, are used for inpainting tasks \cite{Lugmayr2022}. Stable diffusion methods start with a randomly generated two-dimensional noise. Then, this image is denoised iteratively with a previously trained decoder. The approach presented in \cite{Lugmayr2022} uses the incomplete image as input. Noise is added to both the already existing and missing parts. This image is denoised by the diffusion model. After this step, the next iteration starts, but instead of taking noise to inpaint the missing parts, the denoised output of the diffusion model is used for inpainting. Even though these models currently show the best results in the inpainting of natural images, we will not use diffusion models since they rely on repeated application of the decoders, making their use too slow and thus intractable for online experimentation applications.

Other techniques used for inpainting tasks are fully-connected or \ac{CAE} \cite{Givkashi2022}. An autoencoder is a neural network consisting of two parts: The encoder is composed of several layers with decreasing size. The last layer of this encoder is called bottleneck, and its neuron count is called bottleneck size. The second part, the decoder, typically has the same amount of layers but with symmetrically increasing layer sizes. Usually, one trains an autoencoder to reconstruct the input data precisely. This procedure allows us to learn a compressed representation of the input data using the encoder and enables the decoder to unfold the data. Autoencoders provide fast and accurate reconstructions. For autoencoders, extensions exist for taking the circular setup of the \ac{TOF} detectors into account \cite{Schubert2019}. These so-called \acp{CCNN} replace the zero padding from the convolution layers in the encoder by a circular padding. This means that the left-side pixels are padded with values from the right side edge of the input and the right-side pixels are padded with values from the left-side edge. For the transposed convolution layers in the decoder the \acp{CCNN} add padding to their input and cut the outputs than accordingly to their expected output shapes.

As previously described, \acp{GAN} exhibit significant challenges during training, and diffusion models are too slow for our intended online application. Due to this difficulties, fully-connected and convolutional autoencoders---also with the addition of circular padding--- emerge as the most promising approaches. Therefore, this study will focus on evaluating and comparing these autoencoder methods.

To achieve the advances discussed in \cref{sec:introduction_tof}, we need to overcome some largely unexplored challenges. First, we need to investigate ways to transfer knowledge from simulation data to real-world data. Additionally, it is important to avoid overfitting on the simulation, and thus ensure that the resulting model is still general enough to be applied to the real data. Finally, we need to address how we can mitigate the disproportionately negative impact of certain failed \ac{TOF} detectors on overall reconstruction quality.

\section{Method}
\label{sec:tof_rec_method}
\subsection{Simulation \& Dataset}
\begin{center}
\begin{figure*}
    \centering
\includegraphics[width=\textwidth]{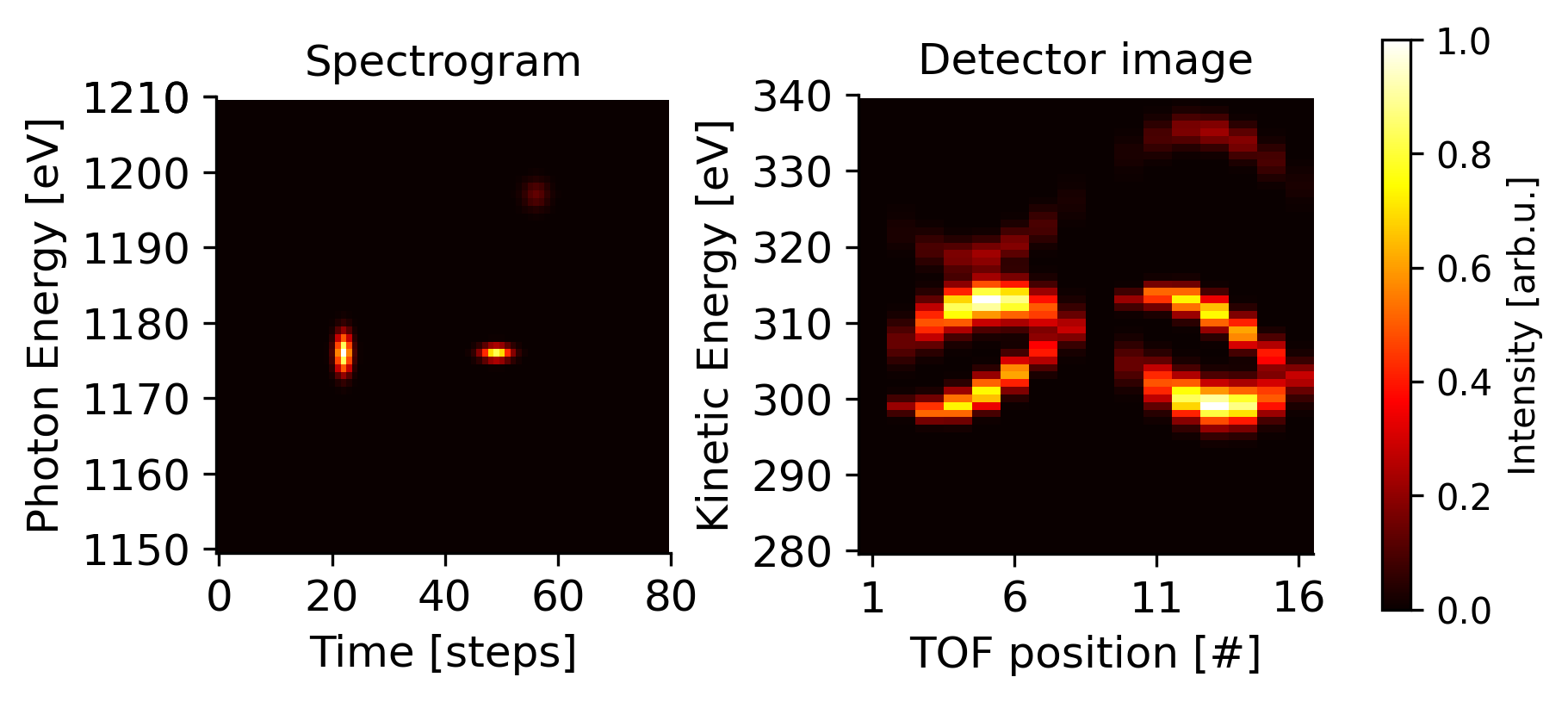}
\caption{
The simulation generates random spectrograms (left) and translates them to the detector images (right).}
\label{fig:spectrogram_detector_image}
\end{figure*}
\end{center}

Using the publicly available (\url{https://github.com/hz-b/tof-reconstruction/blob/main/data_generation.py}) partial-wave-based simulation by \citeauthor{Hartmann2018}, we generate detector images similar to those recorded during the experiments. \Cref{fig:spectrogram_detector_image} shows an example of such a detector image on the right. We will use these detector images later to train the neural networks, because the images generated in this way have the advantage that they are without unknown noise and complete, i.e. the complete data of all TOFs are shown. These detector images consist of 16 \ac{TOF} angles and 60 discrete equidistant kinetic energy steps, respectively. The color shows the intensity, which means the number of electrons per energy level observed at the angularly distributed \ac{TOF} detectors.

The simulation first generates random spectrograms to create these detector images. A spectrogram, as shown on the left side of \cref{fig:spectrogram_detector_image}, represents the time–energy distribution of the X-ray pulses. In \cref{fig:spectrogram_detector_image}, it contains three broad intensity peaks. Even though these spectrograms cannot be measured in real-world experiments, they visually represent the most significant properties of the X-ray pulses. These properties can only be inferred indirectly through the resulting detector images. The generated spectrograms are the sum of 1–5 two-dimensional Gaussian distributions, referred to as peaks in \cref{tab:sim_parameters}. The variance of these distributions is randomly chosen from a uniform distribution $\mathcal{U}\left(0, \sigma_{\text{max}}\right)$, denoted as the number of phase steps, one step equals $\frac{35}{80}$ femtoseconds for the used laser.

The simulation translates this spectrograms to detector images, by imitating the streaking effects of the circularly polarized laser on the photoelectrons in different angular positions due to their emission time. Since this emission is induced by the X-ray pulses, and the circularly polarized laser is synchronized with the X-ray pulses, the peaks from the spectrogram produce the waves on the detector images: The top peak in the spectrogram produces the top wave in the detector image. The left intensity peak at around 20~time~steps produces the broader high intensity wave in the middle on the detector image. The right high intensity peak produces the other wave in the middle of the detector image, showing that their produced waves can also overlap each other. It is also visible that a phase shift (displacement in time axis) results in a displacement of the \ac{TOF} position of the wave and a shift in photon energy results in a displacement in the kinetic energy axis of the detector image. The remaining parameters listed in \cref{tab:sim_parameters} are the settings of the detector setup.

The ellipticity of the polarization laser at some phase step $\phi$ is calculated with the following ellipticity function:
\begin{equation}
    \text{ef}(\phi) = \frac{\varepsilon^2}{\left(\varepsilon \cos(\phi - \theta)\right)^2 + \left(\sin(\phi - \theta)\right)^2}
\end{equation}
The ellipticity $\varepsilon$ describes how closely the polarization of the laser approaches circularity, while the ellipse tilt angle $\theta$ represents the orientation of the polarization ellipse. These values have been measured for this experiment \cite{Hartmann2018}.

\begin{table}[tbp]
\centering
\begin{tabular}{cc}
\hline
parameter & value \\
\hline
peak &  $\left[1, \dots, 5\right]$ \\
$\sigma_{\text{max}}$ & 7~steps \\
kick $\kappa$ & $\left[0, 100\right]$ eV \\
ellipticity $\varepsilon$ & 0.73 \\
ellipse tilt $\theta$ & $\frac{3}{8}\pi$ rad \\
$\beta_2$ & 2 \\
\hline
\end{tabular}
\caption{Parameters and intervals of simulation.}
\label{tab:sim_parameters}
\end{table}

The translation from spectrograms $\mathbf{S}$ to detector images $\mathbf{D}$ works as follows:

\begin{equation}
    d_{ij} = \sum_{i, j=0}^{i=80, j=60}  \text{sim}\left(j, \kappa, \frac{2\pi i}{80}\right) s_{ij}
\end{equation}

The values $d_{ij}$ and $s_{ij}$ are the entries of the according spectrogram $\mathbf{S}$ and the resulting detector image $\mathbf{D}$. The kick $\kappa$ is the amplitude of the waves, i.e. the maximum streaking amplitude of the electrons induced by the streaking laser \cite{Dingel2022}. It is drawn from a uniform distribution $\kappa \sim \mathcal{U}\left(  0, 100  \right) $ for enabling unbiased training, since the kick can vary for every pulse.

The simulation of a partial wave is calculated as follows:
\begin{equation}
\text{sim}\left(e_{\mathrm{kin}}, \kappa, \phi\right) = I_{\mathrm{electron}} \cdot g\left(e_{\mathrm{kin}}, \text{sine}(e_{\mathrm{kin}}, \kappa, \phi), \sigma_E\right)^\top
\end{equation}
We set $\sigma_E = 1 \si{eV}$ as the standardized width of the Gaussian curves, matching the energy step size in the detector images. This choice ensures that any realistic partial waves for the detector image can be constructed.

The angular distribution $\mathrm{I_{electron}}$ is calculated with the following equation \cite{Hartmann2018, Yang1948}:

\begin{equation}
\mathrm{I_{electron}} = \frac{\beta_2}{4} \left(1-\cos\left(2\boldsymbol{\alpha}\right)\right)
\end{equation}

The anisotropy angular distribution parameter $\beta_2$ modulates the shape of the photoelectron angular distribution and is characteristic of the orbital of the target gas. We set $\beta_2=2$ because the s-orbital of neon gas is being excited.

The mathematical representation of all 16 \ac{TOF} detector angles in the dipole plane is denoted with~$\boldsymbol{\alpha}$.

The function $g$ is approximating the Gaussian function with full width half maximum:

\begin{equation}
g\left(x, x_0, x_w\right) = \exp\left(-\frac{(x - x_0)^2}{2 \left(\frac{x_w}{2.35}\right)^2}\right)
\end{equation}

While $x$ is the independent variable for which the Gaussian function is evaluated, $x_0$ is the peak position of the Gaussian curve, $x_w$ is the width of the Gaussian curve at half of its maximum height.

The $\text{sine}$ determines the peak position of the Gaussian curve:
\begin{equation}
\text{sine}\left(e_{\mathrm{kin}}, \kappa, \phi\right) = e_{\mathrm{kin}} + \kappa + \cos \left(\boldsymbol{\alpha} - \phi\right) \text{ef}\left(\phi\right)
\end{equation}
By running the simulation -- generating random spectrograms and translating them to detector images -- 6~million times, the dataset for the experiments in this article is created. It is  stratified by the amount of peaks, thus it contains an equal amount of samples with one to five peaks each. The number of peaks in the dataset is shuffled to prevent bias in the training process. For training, we split this dataset into 4.8~million training and 600\,000 validation samples. Due to the high amount of required evaluations for the following examinations, we limited the test data to 100\,000 samples. Training data is used by the neural network during training, and the validation dataset is used to assess a stopping point for training. The test dataset is not used during training and is utilized for the final evaluation of the trained models. 

\subsection{Noise \& Data Augmentation}
We incorporate different forms of artificial noise into the simulation outputs for our denoising and reconstruction method, aiming to imitate the noise of the actual \ac{TOF} detectors as realistically as possible. Due to the experiment's design, noise levels are unpredictable and can vary significantly across experiments. The detector's noise can be categorized into signal-dependent and signal-independent noise.

The signal-dependent noise increases with the intensity potentially non-linearly and can have electric or physical sources, such as secondary particles or multi-photon effects. Since the variability in noise across different experiments prevents a consistent estimation, we account for this type of noise with high-intensity peak noise in the dataset and training. We model it as a uniformly distributed random variable to avoid assumptions about the noise magnitude.

The signal-independent noise is a white detector noise that is relatively higher for weaker intensities. We address this noise with a Gaussian noise with uniformly distributed variance. Additionally, due to \ac{TOF} detector calibration, saturation effects can be ruled out for our experiment. In \cref{sec:saturation}, showing that the total measured electron intensities by all \ac{TOF} detectors correlate nearly linearly with the total electron intensity measured by a gas monitor detector, rather than transitioning to a flattened curve.

For training, we add noise to the detector images with disabled \ac{TOF} detectors as input and the denoised complete detector image as target. Please note that the order of the following modifications is important. We choose all parameters such that the noise level of the created detector image is visually higher than in the recorded image at the real machine.

\begin{enumerate}
    \item \textbf{Dataset High-Intensity Peaks:}
    During dataset creation, we simulate 1-15 random single pixels of high intensity by adding random, uniformly distributed single pixel spots to the spectrogram. The intensity of these pixels is also random, uniformly distributed. These peaks create isolated, high-intensity spots in the detector images, mimicking spikes as shown in \citeauthor{Hartmann2018} Figure 2c \cite{Hartmann2018}.

    \item \textbf{Training High-Intensity Peaks:} To create this noise, we set a probability $p_{\mathrm{peak}}$ to which every pixel of a \ac{TOF} detector is set to the maximum value of the detector image. We choose the maximum value, so that it imitates a high intensity noise peak. In a later step, we add Gaussian noise to the entire detector image, so that the Gaussian noise is also applied to the high peaks making the simulated detector images more realistic and harder to detect by the denoising mechanism. For the high-intensity peak noise, we generate a matrix $\mathbf{P}$ with equally distributed random values as shown in the following equation:

\begin{equation} 
\label{eq:hot_peak_distribution}
    p_{ij} \sim \mathcal{U}\left(0, 1\right).
\end{equation}

This matrix has the same size as the detector images $i \times j$, where $i \in \left\{ 1, \dots, 16\right\}$ is the \ac{TOF} detector position and $j \in \left\{1, \dots, 60 \right\}$ is the amount of energy steps, which corresponds to the resolution of the \ac{TOF} detectors.

If an entry $p_{ij}$ is less or equal to $p_{\mathrm{peak}}$, it is set to the maximum value of the current detector image as follows:
\begin{equation}
\label{eq:hot_peak_calculation}
\bar{x}_{ij} = \begin{cases}
 x_{ij}, & \text{if} \quad p_{ij} \leq p_{\mathrm{peak}} \\
 \max_{kl}{(x_{kl}),} & \text{else.}
\end{cases}
\end{equation}

\item \textbf{Normalization:} Next, we min-max normalize the detector images. We apply the normalization individually per image. The real-world images are also normalized per image since only the relative differences per image should be considered, given the technical functionality of \ac{TOF} detectors.

\item \textbf{Gaussian Noise:} It is important that the previous normalization step is performed first, ensuring that the Gaussian noise is added to the normalized values, as illustrated in the following equation:
\begin{equation}
\tilde{x}_{ij} = \bar{x}_{ij} + p_{ij}
\end{equation}

The variables $i$ and $j$ are chosen similar to \cref{eq:hot_peak_distribution}. The entries of the random matrix $\mathbf{P}$ are drawn from a Gaussian distribution:

\begin{equation}
p_{ij} \sim \mathcal{N}\left(0, \sigma^2\right)
\end{equation}

The noise level can be chosen by setting $\sigma$ which is the standard deviation of the Gaussian distribution. We choose a random noise level of $\sigma \sim \mathcal{U}\left(0, 0.2\right)$, so that the noised images look as similar as possible to the real images.

\item  \textbf{Detector disablement}: The next step is to disable the \ac{TOF} detectors, with specific methods applied for training each model, as listed in \cref{tab:models}. In the following, '\#' denotes a \ac{TOF} detector's position, while a number before \ac{TOF} detectors indicates the count of uniformly random-selected \ac{TOF} detectors. The selection strategy of which \ac{TOF} detectors to disable is shown in detail in \cref{alg:random_tofs}. The first \ac{TOF} detector is selected randomly. All others are picked randomly as well, or, with probability $\xi$, a neighboring or opposite \ac{TOF} detector is chosen for disabling if possible.

\begin{table}
    \centering
    \begin{tabular}{ll}
    \hline
        Model & Training Data \\
        \hline
        1\ac{TOF}, 2\ac{TOF}, 3\ac{TOF} & Disable 1, 2 or 3 random \ac{TOF} detectors respectively \\
        General & Disable 1–3 \ac{TOF} detectors randomly\\
        Spec & Disable the \ac{TOF} detectors at positions \#8 and \#13\\
        Mean & No data; Calculates means only from neighboring \ac{TOF} detectors \\
        \hline
    \end{tabular}
    \caption{Different models trained with varying datasets adapted for specific purposes.}
    \label{tab:models}
\end{table}

\item \textbf{Normalization:} Min-max normalization is applied to each image individually again, ensuring that the values of the detector images---captured with disabled \ac{TOF} detectors---are rescaled between zero and one. This normalization process can also be implemented in real-world scenarios where \ac{TOF} detector failures occur. This approach remains valid since we focus on the time-energy structure rather than the beam intensity. Furthermore, as demonstrated in \cref{sec:saturation}, the X-ray photon energies can easily be extracted from the raw data and correlate almost linear with the corresponding values of the gas monitor detector. This correlation enables reconversion of the detector images to their original photon intensity level and thus ensuring no loss of essential information.

In \cref{sec:saturation}, showing that the total measured electron intensities by all \ac{TOF} detectors correlate nearly linearly with the total electron intensity measured by a gas monitor detector, rather than transitioning to a flattened curve.

\end{enumerate}

As a baseline, we will use the \textit{mean} model: This model averages over the neighbors' values to replace the values of a missing \ac{TOF} detector. If a neighbor of a missing \ac{TOF} detector is also missing, we take the nearest neighbor. Due to the experimental setup's circular nature, every \ac{TOF} detector has two neighbors since the first \ac{TOF} detector is located next to the last. Unlike the other models based on simulation data, this model is entirely data-free, meaning it relies only on information from the neighboring \ac{TOF} detectors in the current detector image without requiring any training data or pre-computed statistics. By checking if our proposed neural network models have a lower loss value than the mean model, we verify that the neural network models do not only learn to mean over the neighbors of a missing \ac{TOF} detector.

\subsection{Models}
For the neural network for denoising and \ac{TOF} detector reconstruction, we assess several neural network architectures:
We train a \ac{CAE} architecture with 5 hidden layers, which we refer to as \ac{CAE}-64, with filter sizes 32, 64, 128, 256, 512, and in the bottleneck 64 for the encoder, similarly for the decoder but in reverse order. The filter has dimensions $3\times3$, which is small enough to capture details but large enough to significantly reduce spatial dimensions. The stride is set to 2 to further reduce spatial dimensions. The padding is set to 1 to avoid underrepresentation of the border pixels. The output padding is set to 1 for a better fit to the encoder dimensions. Finally, we add a bilinear upsampling layer for getting precisely the desired output shape of $60\times16$, matching the simulation outputs: 16~\ac{TOF} angles and 60~discrete, equidistant kinetic energy steps.

In addition, we assess alternative architectures with smaller or larger bottleneck dimensionality to provide the possibility of increasing the denoising level and decreasing reconstruction details. These architectures have filter sizes of 32, 128, 256, 512, and 1024 in the bottleneck. They replace the bottleneck layer with 64 filters from the \ac{CAE}-64. This design choice reduces redundancy and inefficiency while mitigating the risk of vanishing gradients, which can occur when no significant transformations are performed. We refer to these architectures as \ac{CAE}-32, \ac{CAE}-128, \ac{CAE}-256, \ac{CAE}-512, and \ac{CAE}-1024.

Due to the circular arrangement of the \ac{TOF} detectors, the \ac{TOF} detector at the most right in the simulation is adjacent to the most left \ac{TOF} detector. We test the \acp{CCNN} described in \cref{sec:tof_rel_work} to model this occurrence. This extension replaces the zero padding in the convolution layers with circular padding. In circular padding, the left-side pixels are padded with values from the right-side edge of the input, and the right-side pixels are padded with values from the left-side edge.

We also tested U-Nets as presented in \cite{Hosen2022} with a similar structure as the \ac{CNN} autoencoder architecture, but since these networks were overfitting quickly, we did not further investigate this approach.

As an activation function, the \textit{Mish} function proposed in \cite{Misra2020} outperformed the frequently used ReLU function. Furthermore, we used the AdamW optimizer as presented in \cite{Loshchilov2019}. The regularization term of the optimizer results in increased training and validation losses. However, the models denoise and reconstruct visually more accurately on real-world images, as shown in \cref{ssec:regularization_impact}. Thus, AdamW provides a countermeasure against overfitting on the simulation. We use learning rate plateau scheduling, beginning with a learning rate $\eta=10^4$ and decaying it exponentially every epoch by multiplying it with $0.1$ if the validation loss has not improved more than $\epsilon = 10^{-8}$ for three epochs. Despite the outputs being min-max normalized and thus between 0 and 1, using no activation function after the last layer resulted in slightly more accurate reconstruction and denoising than using a Sigmoid output function. We stopped the training after 50 epochs, since no significant increase in validation loss was noticeable. Using these best neural network parameters and architecture, we train the different models \textit{1TOF}, \textit{2TOF}, \textit{3TOF}, \textit{general} and \textit{spec} with differently augmented data as listed in \cref{tab:models} and compare those for different settings.

As baseline for the denoising process, we use a Wiener filter as described by \citeauthor{Lim1990} \cite{Lim1990}. This Wiener filter works by minimizing the \ac{MSE} between the estimated output signal and the original signal. It adapts its response based on both the signal and noise characteristics. We use a $3\times3$ filter window size, following our choice in the \ac{CAE}. The noise is estimated as the average of the local variance -- that refers to the variance in the corresponding filter window -- of the input. The denoising process with Wiener filter is described more detailed in \cref{sec:wiener}.

\subsection{Real-World Data Application}
We use detector images from an experiment at \ac{LCLS} as presented in \cite{Hartmann2018} to show our method in a real-world setting. The time of flight $t$ measured by the \ac{TOF} spectrometers correlates with the energy $E$ roughly proportional to $\text{const} + \frac{1}{t^2}$. Measurements are taken equidistantly in time of flight and then mapped onto equally sized energy intervals. Consequently, the number of measurement points in time of flight varies per equidistant energy $E$ interval. This mapping is used because it is physically easier to measure equal intervals in terms of time of flight rather than energy. Furthermore, the scaling factors differ for all \ac{TOF} detectors, and it would be required to remeasure and calculate the time-to-energy conversion for each experiment. Thus, it is easier to map the energies to equidistant 1~eV bins rather than adapting the simulation for every experiment. 

Because of this binning of the energy values in 1~eV steps in the simulation, some bins always remain empty since, in the area with lower resolution, no \ac{TOF} detector measures in the specific 1~eV range. To mitigate this issue, if a bin stays empty, we distribute a third of the values of the two neighboring bins to the empty bin. If an empty bin has only one neighbor, we shift half of the value of this bin to the empty one.

Furthermore, we round the positions of the \ac{TOF} detectors scientifically. For example, the 280.5~eV \ac{TOF} detector value is added to the 281~eV bin. After this procedure, we shift all negative \ac{TOF} detector values to $0.0$ and min-max normalize every detector image.
\section{Evaluation}
\begin{center}
\begin{figure*}
    \centering
\includegraphics[width=\textwidth]{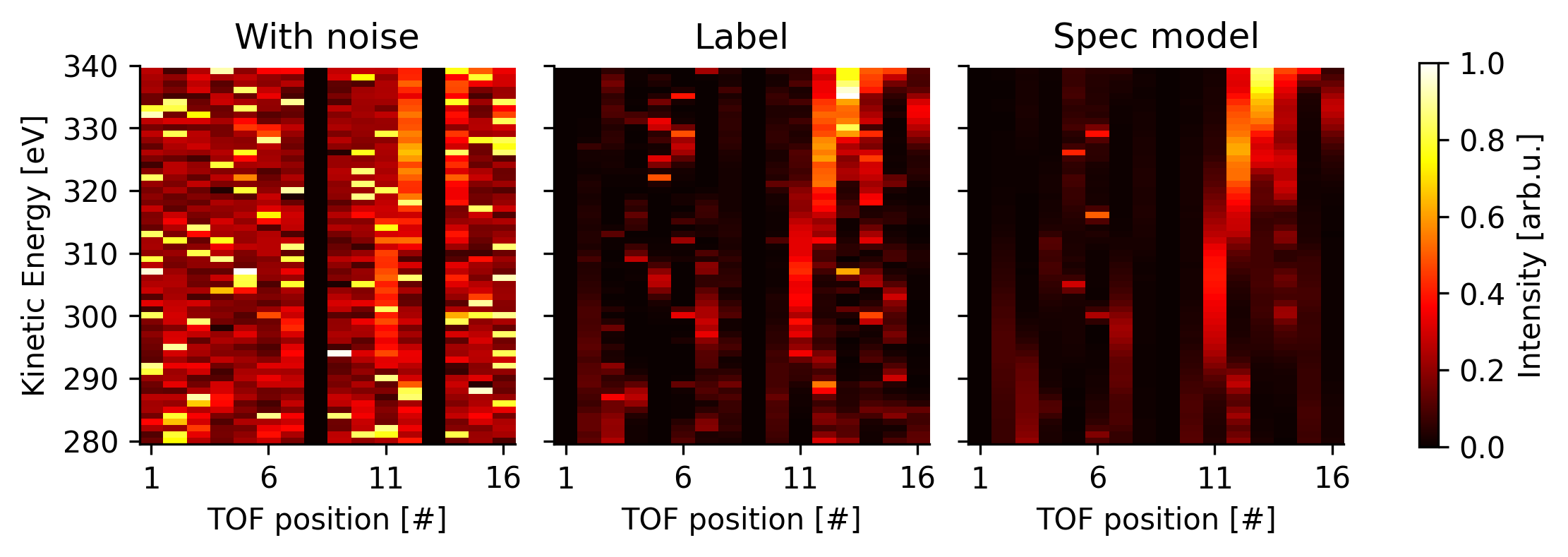}
\caption{
The sample on the left is with two disabled \ac{TOF} detectors, Gaussian noise and hot peaks. In the middle is the same sample without disabled \ac{TOF} detectors and noise. The right plot shows the reconstructed and denoised detector image.}
\label{fig:two_tofs_disabled}
\end{figure*}
\end{center}

\begin{table}[tbp]
            \centering
\begin{tabular}{p{2.5cm}|c}
       \hline
Failed \acp{TOF} & 0 \\
scenario & random \\
\hline
1TOF model & 2.48e-3 $\dagger$ \\
2TOF model & 3.10e-3 $\dagger$ \\
3TOF model & 4.34e-3 $\dagger$ \\
General model & \textbf{2.37e-3} \\
Spec model & 5.04e-3 $\dagger$ \\
Mean model & 7.80e-2 $\dagger$ \\
\hline
\end{tabular}

\caption{\ac{RMSE} on simulated data without failed \ac{TOF} detectors for all models, showing model performance on only denoising.}

\label{tab:evaluation_models_0_failed}
        \end{table}

\begin{table}[tbp]
            \centering
\begin{tabularx}{\textwidth}{p{2.5cm}|*{7}{>{\centering\arraybackslash}X}}
        \hline
Failed \acp{TOF} & 1 & 2 & 3 & 1--3 & 2 & 2 & \#8,\#13 \\
scenario & random & random & random & random & neighbors & opposite & position \\
\hline
CAE-32 & 3.27e-3 $\dagger$ & 3.42e-3 $\dagger$ & 3.64e-3 $\dagger$ & 3.47e-3 $\dagger$ & 3.50e-3 $\dagger$ & 3.37e-3 $\dagger$ & 3.40e-3 $\dagger$ \\
CAE-64 & 2.48e-3 $\dagger$ & 2.64e-3 $\dagger$ & 2.85e-3 $\dagger$ & 2.68e-3 $\dagger$ & 2.72e-3 $\dagger$ & 2.60e-3 $\dagger$ & 2.60e-3 $\dagger$ \\
CAE-128 & 1.93e-3 $\dagger$ & 2.10e-3 $\dagger$ & 2.33e-3 $\dagger$ & 2.14e-3 $\dagger$ & 2.18e-3 $\dagger$ & 2.06e-3 $\dagger$ & 2.06e-3 $\dagger$ \\
CAE-256 & 1.54e-3 $\dagger$ & 1.73e-3 $\dagger$ & 1.97e-3 $\dagger$ & 1.77e-3 $\dagger$ & 1.80e-3 $\dagger$ & 1.69e-3 $\dagger$ & 1.72e-3 $\dagger$ \\
CAE-512 & \textbf{1.40e-3} & \textbf{1.63e-3} & \textbf{1.93e-3} & \textbf{1.68e-3} & \textbf{1.72e-3} & \textbf{1.59e-3} & \textbf{1.62e-3} \\
Mean model & 7.77e-2 $\dagger$ & 7.76e-2 $\dagger$ & 7.76e-2 $\dagger$ & 7.77e-2 $\dagger$ & 7.81e-2 $\dagger$ & 7.72e-2 $\dagger$ & 7.77e-2 $\dagger$ \\
\hline
\end{tabularx}
\caption{\ac{RMSE} comparison across different architectures for different failure scenarios, with 1 to 3 \ac{TOF} detectors disabled.}
\label{tab:reconstruction_architectures}
\end{table}

\begin{table}[tbp]
            \centering
        \begin{tabularx}{\textwidth}{p{2.5cm}|*{7}{>{\centering\arraybackslash}X}}
        \hline
Failed \acp{TOF} & 1 & 2 & 3 & 1--3 & 2 & 2 & \#8,\#13 \\
scenario & random & random & random & random & neighbors & opposite & position \\
\hline
1TOF model & 2.48e-3 & 4.00e-3 $\dagger$ & 6.61e-3 $\dagger$ & 5.06e-3 $\dagger$ & 6.72e-3 $\dagger$ & 2.76e-3 $\dagger$ & 2.74e-3 $\dagger$ \\
2TOF model & 2.64e-3 $\dagger$ & 2.67e-3 $\dagger$ & 3.15e-3 $\dagger$ & 2.90e-3 $\dagger$ & 2.76e-3 $\dagger$ & 2.63e-3 $\dagger$ & 2.63e-3 $\dagger$ \\
3TOF model & 3.17e-3 $\dagger$ & 2.80e-3 $\dagger$ & 2.86e-3 & 2.97e-3 $\dagger$ & 2.87e-3 $\dagger$ & 2.78e-3 $\dagger$ & 2.74e-3 $\dagger$ \\
General model & \textbf{2.48e-3} & \textbf{2.64e-3} & \textbf{2.86e-3} & \textbf{2.69e-3} & \textbf{2.72e-3} & \textbf{2.60e-3} & 2.60e-3 $\dagger$ \\
Spec model & 9.99e-3 $\dagger$ & 1.36e-2 $\dagger$ & 1.68e-2 $\dagger$ & 1.34e-2 $\dagger$ & 1.35e-2 $\dagger$ & 1.34e-2 $\dagger$ & \textbf{2.40e-3} \\
Mean model & 7.76e-2 $\dagger$ & 7.75e-2 $\dagger$ & 7.76e-2 $\dagger$ & 7.77e-2 $\dagger$ & 7.81e-2 $\dagger$ & 7.73e-2 $\dagger$ & 7.78e-2 $\dagger$ \\
\hline
\end{tabularx}

\caption{\ac{RMSE} for different failure scenarios, where 1 to 3 \ac{TOF} detectors are disabled.}
\label{tab:evaluation_models_vs_failed}
\end{table}

In this section, we will first evaluate the denoising performance on detector images of our models. After that, we will examine the reconstruction of failed \ac{TOF} detectors. For both applications, we will provide and discuss a real-world example.

\subsection{Denoising}
As shown in \cref{tab:evaluation_models_0_failed}, we compare the \ac{RMSE} of different models. The scenario \textit{0 failed} means that no \ac{TOF} detector is disabled, thus the models perform only denoising. It is noticable that the specifically trained models are worse in this scenario than the General model. Compared to the respectively best model, the baseline mean model is worse by almost a factor of 10 in terms of the \ac{RMSE}. The $\dagger$ next to an \ac{RMSE} value indicates a significant difference from the best models' \ac{RMSE}. To assess whether differences in the RMSE values are meaningful, we compare the 99\% \acp{CI} of the means and consider them significantly different if the \acp{CI} of this difference does not contain zero. Given the large test sample size of 600\,000, we provide the \acp{CI} in \cref{tab:denoising_statistics} to assess whether the statistically significant differences are also practically relevant. Since the \acp{CI} are in the range of the \ac{RMSE} values, we can consider the differences between the compared models and the Spec model practically relevant.

In \cref{fig:real_image} we show a comparison of the original real data and denoised with the model only trained on the scenario of failed \ac{TOF} detectors \#8 and \#13. In \cref{fig:real_image_disabled_2_tofs} we deleted the data for two \ac{TOF} detectors at positions \#8 and \#13 and used the model specifically trained to reconstruct the \ac{TOF} detectors \#8 and \#13. We decided to show the results on this combination, since it contains one easier and one harder \ac{TOF} detector to reconstruct, according to \cref{fig:1_tof_failed}. We can see that the image, the model produces, is similar to the reconstruction of the detector image with all \ac{TOF} detectors.

The \ac{RMSE} between the Wiener filter output and the \ac{CAE}-64 output is 0.143, while the \ac{RMSE} between the original image and the \ac{CAE}-64 output is 0.150. Additionally, the \ac{RMSE} between the original images and the Wiener filter outputs is 0.072, which is significantly smaller compared to the differences with the \ac{CAE}-64 outputs.

This shows that the \ac{CAE}-64 output images are closer to the Wiener-filtered images than to the original images. However, the changes introduced by the Wiener filter are much smaller than those introduced by the \ac{CAE}-64.

\subsection{Reconstruction}
In \cref{fig:two_tofs_disabled} we show an example of a noisy image from simulation with two disabled \ac{TOF} detectors in the image on the left. We used the Spec model for denoising and reconstruction of the missing \ac{TOF} detectors shown on the right image. The Spec model reconstructs most of the details in the label (center), while providing a fitting reconstruction of the missing \ac{TOF} detectors. In \cref{tab:reconstruction_architectures} we show the different architectures in terms of different sized bottlenecks. There is a clear connection between the bottleneck size and the reconstruction quality on synthetic data. The comparison of architectures with varying bottleneck sizes shows that larger bottlenecks result in more minor reconstruction errors in simulation data. The \ac{CAE}-512 outperforms all other models in terms of the reached \ac{RMSE}. 

However, this behavior cannot be transferred to the real-world images: Larger bottlenecks exhibit weaker denoising performance when applied to real-world data, and the reconstructed \ac{TOF} detector data integrate less cohesively with the rest of the detector image, as shown in \cref{fig:real_image_disabled_2_tofs}. Additionally, larger bottlenecks tend to reproduce noise peaks in the output. Conversely, smaller bottlenecks yield detector images closer in appearance to simulation data while being less detailed than the original real-world detector images. The outputs of the \ac{CAE}-32 architecture tend to be overly simplistic, while larger bottlenecks allow for too many details, including noise artifacts. The \ac{CAE}-64 balances reconstruction accuracy and effective denoising, making it the focus of this study. Nevertheless, other architectures may prove beneficial for different application scenarios.

\begin{center}
\begin{figure*}
    \centering
\includegraphics[width=\textwidth]{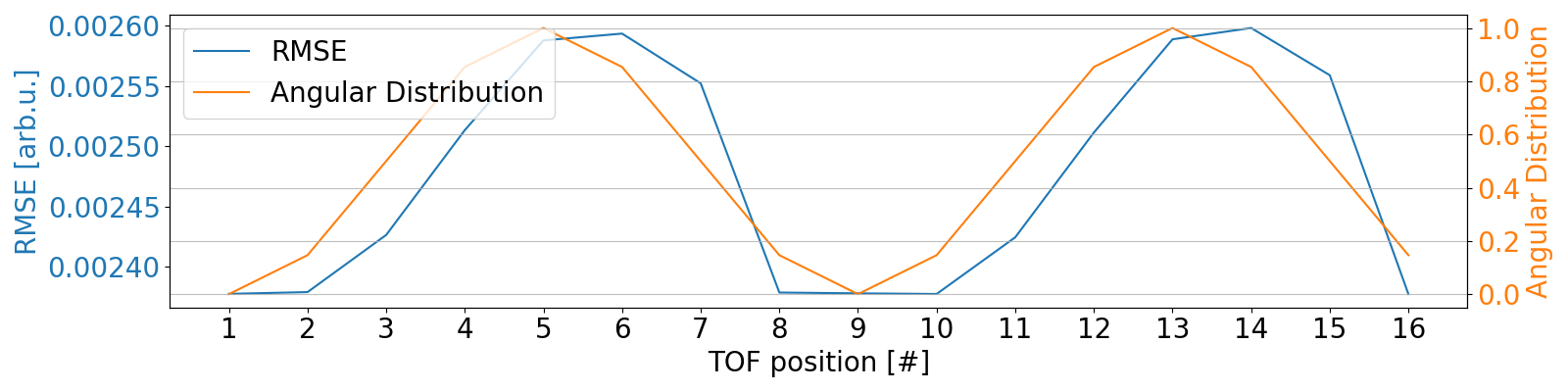}
\caption{
The \ac{RMSE} values of the reconstruction if one \ac{TOF} detector has failed (blue). The y-axis is truncated for emphasizing the differences. For reference, we also plotted the angular distribution (orange).}
\label{fig:1_tof_failed}
\end{figure*}
\end{center}

In \cref{tab:evaluation_models_vs_failed}, we quantitatively compare the models 1TOF, 2TOF, 3TOF with the general and mean model for different scenarios. In the \textit{random} scenario, the number of disabled \ac{TOF} detectors is drawn randomly from a uniform distribution. In the \textit{neighbors} scenario, failures are introduced by disabling \ac{TOF} detectors that are adjacent to each other. In the \textit{opposite} scenario, opposing \ac{TOF} detectors are disabled whenever possible. Finally, in the \textit{position} scenario, \ac{TOF} detectors at positions \#8 and \#13 are consistently disabled. The according general model performs significantly better than all other models, even if the model is trained for a more specialized scenario. The only exception is the 1TOF model for the scenario of one random \ac{TOF} detector failure, but this difference is not statistically significant. The only exception is the scenario of the failure of \ac{TOF} detectors at position \#8 and \#13, where the Specific model that is trained for failure of these specific positions performs significantly better. In \cref{tab:failure_statistics} we provide additional statistics with the \acp{CI}. The mean model is in all cases almost worse by a factor of 20 in comparison to the best model. In \cref{fig:1_tof_failed} we show the evaluation of the test dataset on the General model, i.e. 600\,000 simulated examples. We evaluate the model with one missing \ac{TOF} detector at all 16 possible positions and calculate the mean \ac{RMSE}. We can see that the \ac{RMSE} values of different \ac{TOF} detector positions are pretty similar, but have slight peaks at position \#6 and \#14. This behavior arises because the angular distribution resulting from the chosen target gas leads to reduced intensity in this region, making it more challenging to predict. In \cref{ssec:phase_evaluation}, we provide an additional, phase-separated examination of a single failed \ac{TOF} detector.

\begin{center}
\begin{figure*}
    \centering
\includegraphics[width=0.8\textwidth]{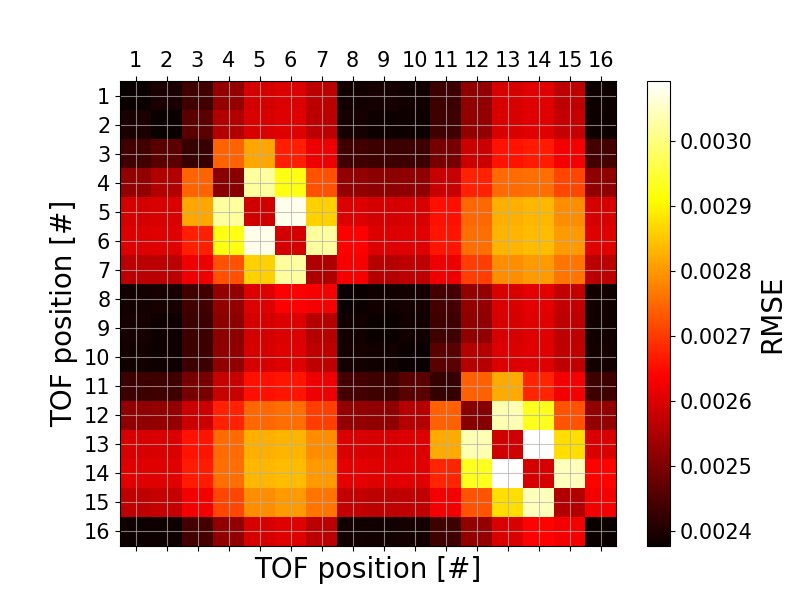}
\caption{
The \ac{RMSE} values of the reconstruction if two \ac{TOF} detectors have failed. }
\label{fig:2_tof_failed}
\end{figure*}
\end{center}

\begin{center}
\begin{figure*}
    \centering
\includegraphics[width=1.0\textwidth]{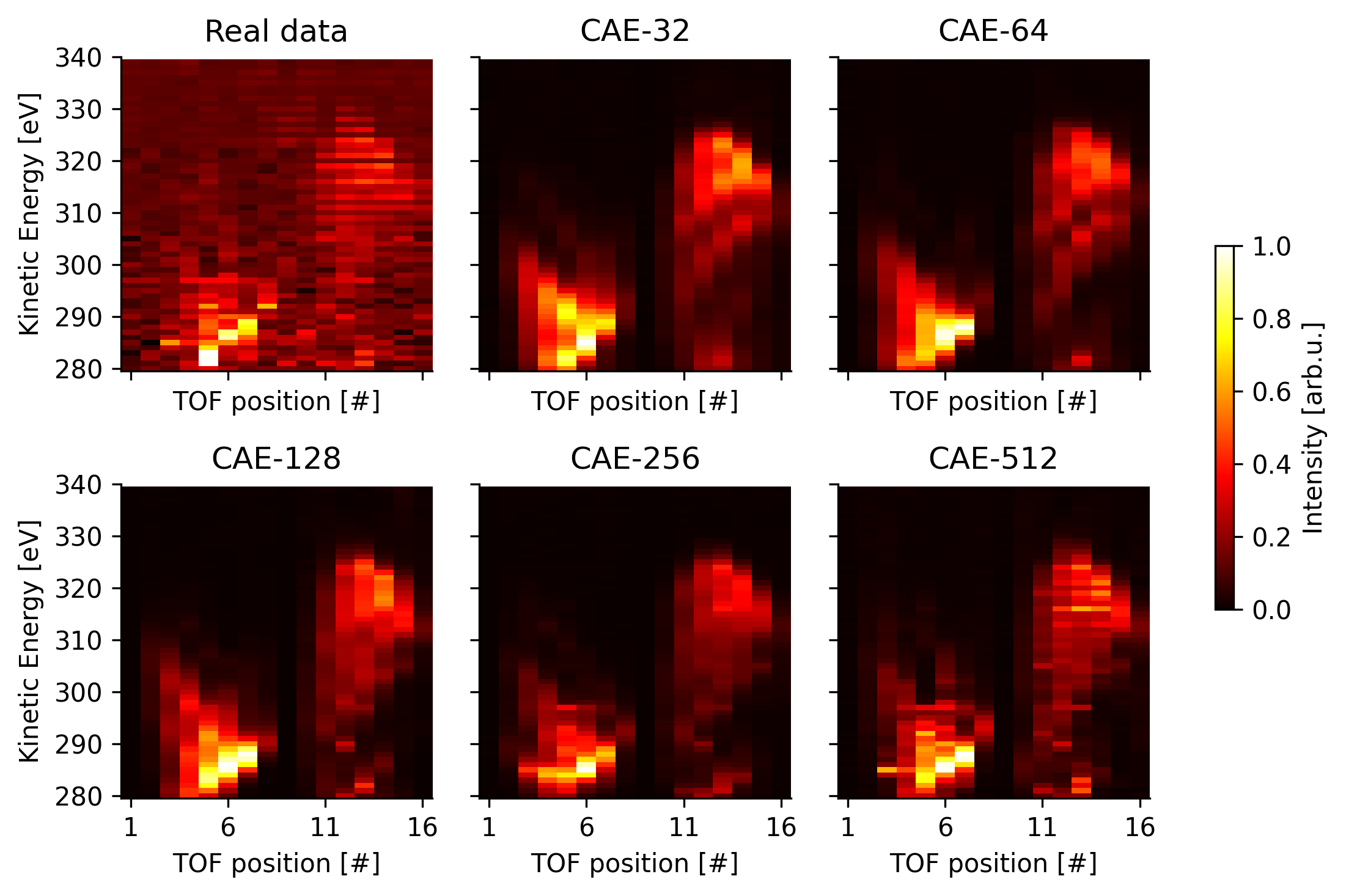}
\caption{
The first plot shows a real-world detector image. The following plots show the denoised detector images across different-sized bottleneck model architectures. All models in these plots are trained in the general setting, that means with 1--3 \ac{TOF} detectors disabled.}
\label{fig:real_image}
\end{figure*}
\end{center}

\begin{center}
\begin{figure*}
    \centering
\includegraphics[width=1.0\textwidth]{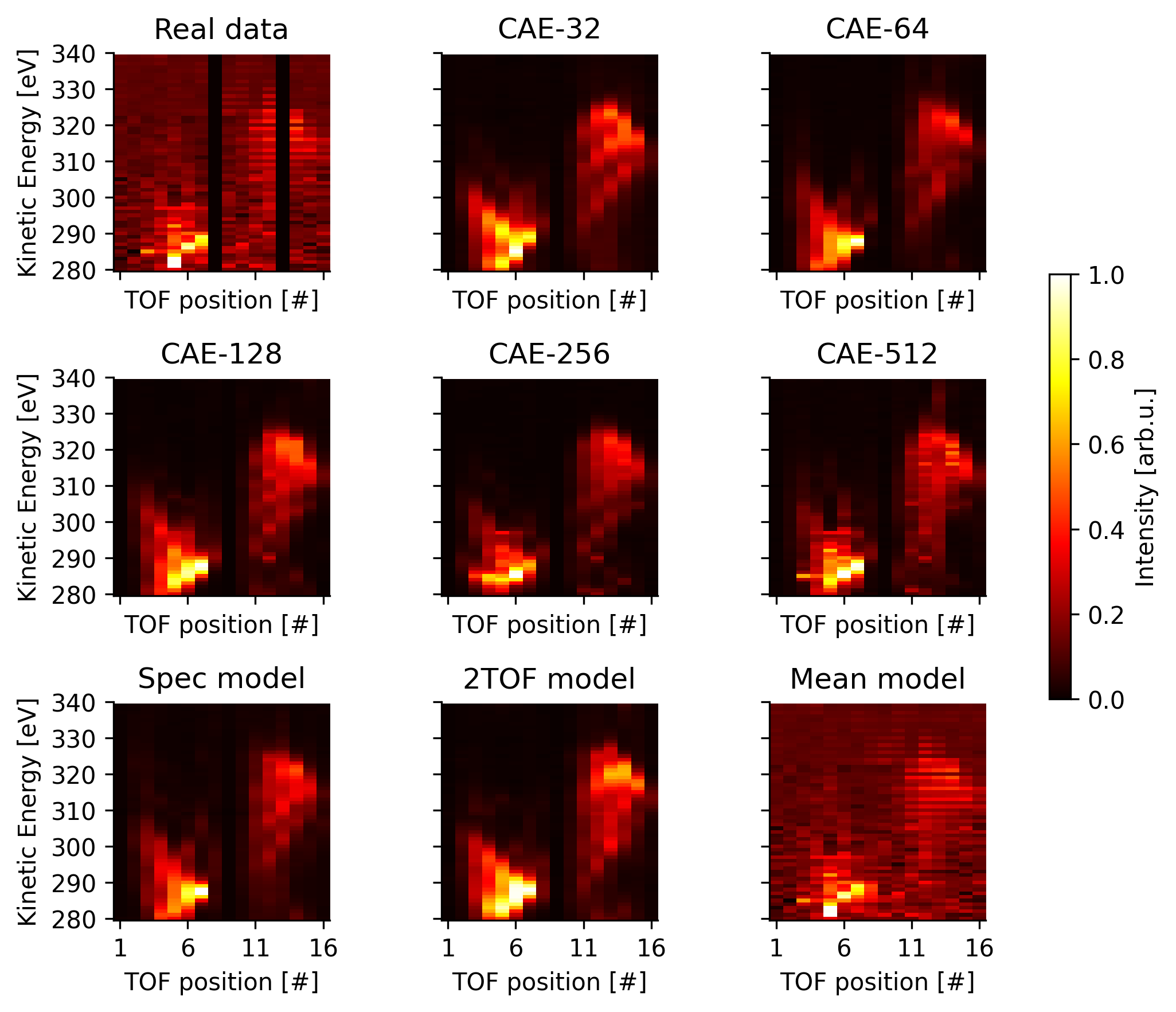}
\caption{
The first plot shows a real-world detector image with two \ac{TOF} disabled. The following plots show the denoised and reconstructed detector images across different model architectures. All models in these plots are trained in the general setting, that means with 1--3 \ac{TOF} detectors disabled, except for the three models in the bottom line: The first one is the output of the \ac{CAE}-64 model trained with data in spec setting (\ac{TOF} positions \#8 and \#13 disabled, exactly the positions disabled for this reconstruction). The second one is the \ac{CAE}-64 model trained with always two \ac{TOF} detectors disabled. The last one is the output of the Mean model, that recreates the missing \ac{TOF} sensors by averaging over its neighbors.}
\label{fig:real_image_disabled_2_tofs}
\end{figure*}
\end{center}

In \cref{fig:2_tof_failed} we show a matrix of the evaluation of \ac{RMSE} values for two failed \ac{TOF} detectors. If a value is on the diagonal, we plot the \ac{RMSE} value of the particular single failed \ac{TOF} detector for reference. The darker the plot's color, the lower the occurring \ac{RMSE}. The highest error occurs if \ac{TOF} detectors \#13 and \#14 fail, followed by \#5 and \#6 in a repeating pattern shifted by 8 \ac{TOF} detectors. Generally, the combinations with \ac{TOF} positions \#5, \#6, \#13, and \#14 any other failed \ac{TOF} detector have higher errors than the rest. This behavior also stems from the angular distribution of the experimental setup, similar to the single failed \ac{TOF} detector examination. The best reconstructions are possible for combinations with \ac{TOF} position \#1, \#2, \#8, \#9 and \#16 have failed. This behavior can be explained since these \acp{TOF} detectors receive no electrons or only negligible amounts. Thus, these parts of the detector image can be reconstructed with smaller errors. The occurring patterns in reconstruction errors if two \ac{TOF} detectors have failed fit to the patterns when one \ac{TOF} detector has failed, as can be seen on the diagonal entries.

\section{Results \& Discussion}
The evaluation in the previous section has shown that the models are able to accurately denoise and reconstruct 1--3 failed \ac{TOF} detectors. We also tested the reconstruction of higher \ac{TOF} detector counts, however the reconstruction quality decreases rapidly for 4 or 5 failed \ac{TOF} detectors, as shown in \cref{tab:failure_statistics_4_5}. We evaluated our models for several scenarios on an extensive test set that was not present during the training phase of the models.

The General model shows the best reconstruction quality if no \ac{TOF} detectors have failed. This behavior is not surprising since no model was trained for that specific scenario, and due to its generality, this model is the most robust and thus can handle unseen cases the best. All presented neural network models reconstruct the missing \ac{TOF} detectors considerably better than the baseline mean model. Furthermore, we have shown qualitatively that our models work on real-world images. The produced denoised detector images look similar to the noisy real-world images. If we additionally turn off \ac{TOF} detectors on the real-world images, the model produces a similar image to a complete input detector image. A numerical evaluation for real-world images is impossible since we cannot generate noiseless detector images for comparison in a real-world scenario.

The comparison of the failure of one specific \ac{TOF} detector shows that the position of a \ac{TOF} has a considerable impact on its reconstruction loss, as shown in \cref{fig:1_tof_failed}. A higher value in the angular distribution -- that is due to the choice of the target gas and the experiment setup -- means these areas often have lower intensities and more possible shapes. Also, combinations of neighbors and opposite \ac{TOF} detectors result in the higher intensity regions of \ac{TOF} detector positions \#4-\#7 and positions \#12-\#15 a slight increase in reconstruction error. This phenomenon can be explained because neighboring or opposite \ac{TOF} detectors share more information than other combinations. If both \ac{TOF} detectors fail, more information is lost compared to other combinations. In the lower intensity regions of \ac{TOF} detector positions \#1-\#3 and positions \#8-\#11 the differences in the reconstruction \ac{RMSE} values are considerably less present. These aspects could be considered in future designs of angular streaking experiments.

Regarding neural network training, one of the essential things is regularization; we use the AdamW optimizer as described in \cite{Loshchilov2019}. This technique helps to keep the weights of the neural network low and thus avoids overfitting not only on training data but also on the simulation.

In \cref{tab:reconstruction_parameters_statistics} we show the evaluation of different neighboring factors $\gamma$. The ratio of shown training samples does not significantly change the training behavior since the \ac{RMSE} is almost equal for all scenarios. Since there is no adverse effect, we set $\gamma=0.5$ for all subsequent experiments. We also tested the previously described \acp{CCNN}, which even decreased the reconstruction error. This behavior could be evoked by the cropping that is required in the transposed convolution layers in the decoder due to the additional paddings in the encoder part. Also the addition of padding to the input data decreases the reached \acp{RMSE}, this might happen because additional input comes to the cost of additional neurons that need to be learned.

The comparisons of the different models, the General model for 1--3 failed \ac{TOF} detectors, and the models for the distinct number of failed \ac{TOF} detectors show that the more specialized a model is, the more accurate its reconstructions are. However, we can verify this only for the simulated test data, even though the visual comparisons indicate similar results for the real-world samples. This behavior is probably caused by the more limited ranges of restored intensities in the case of more specialized scenarios. Thus, the trained model's predictions are more accurate.

\section{Conclusion \& Outlook}
The presented method provides online denoising and reconstruction of missing \ac{TOF} detectors. We have shown that our presented approach for reconstructing missing \ac{TOF} detectors outperforms a simple algorithm that calculates the mean over the remaining neighbors of a missing \ac{TOF} detector. Validating our approach on real-world detector images is difficult because no labels exist. However, the results are visually viable, and due to its fast inference time, our presented method can be used during experimentation time for immediate monitoring and extraction of attributes for further pulse characterization. A first step towards this has been shown in \cite{Dingel2022}: Extracting the kick, Auger decay time, pulse structure, and duration is possible.
The pulse structure is the intensity of a pulse over time, and the pulse duration is the period between a pulse's first and last intensity. Complete and denoised data are inevitably required to obtain correct characterization results.

Alternative layouts for the angular streaking experiment can be efficiently tested and evaluated through simulation to optimize the design of experimental setups. These setups include configurations with higher overlapping \ac{TOF} detector layouts or smaller \ac{TOF} detectors, which correspond to smaller solid angles. A smaller solid angle refers to a reduced portion of the detector's field of view covered by each \ac{TOF} detector, this can be compensated by increasing the number of detectors. Also, to test other scenarios or setups, the other simulation parameters can be changed, such as peak numbers or kick ranges. A realistic amount of noise can be added for all these scenarios, and the expected number of failed \ac{TOF} detectors can be wiped and reconstructed. The error of denoising and reconstruction indicates if there is enough redundant information encoded so the detector image can be reconstructed.

The recently emerged diffusion models could be tested to improve our approach's accuracy and robustness. However, this type of network typically requires at least 20 iterations for inference, which negatively impacts inference time, making them impracticable for an online evaluation setting. Nevertheless, a more accurate offline evaluation could benefit from applying diffusion models.

\begin{acknowledgments}
Support by the JointLab AIM-ED between Helmholtz-Zentrum f\"ur Materialien und Energie, Berlin and the University of Kassel is gratefully acknowledged. We also acknowledge the useful feedback from the co-authors, Shang Gao and Pascal Plettenberg.
\end{acknowledgments}

\appendix
\section{Evaluation Statistics}
In this section we list the detailed statistics of the evaluated denoising and reconstruction scenarios for different models. The first number is the mean, followed by the standard deviation. In braces we supply the 99\% \acp{CI} of the related t-test to check whether the differences of the \acp{RMSE} achieved by a model are statistically significant in comparison to the best evaluated model.
\begin{table}[ht]
            \centering      
        \begin{tabular}{p{2.5cm}|c}
        \hline
Failed \acp{TOF} & 0 \\
scenario & random \\
\hline
1TOF model & 2.48e-3 $\pm$2.08e-3 (-1.08e-4, -9.84e-5) \\
\hline
2TOF model & 3.10e-3 $\pm$3.18e-3 (-7.42e-4, -7.05e-4) \\
\hline
3TOF model & 4.34e-3 $\pm$1.12e-2 (-2.06e-3, -1.88e-3) \\
\hline
General model & \textbf{2.37e-3 $\pm$2.06e-3} \\
\hline
Spec model & 5.04e-3 $\pm$2.74e-3 (-2.68e-3, -2.65e-3) \\
\hline
Mean model & 7.80e-2 $\pm$1.23e-2 (-7.57e-2, -7.55e-2) \\
\hline
\end{tabular}
       
\caption{Statistics for denoising scenario, where no \ac{TOF} detectors are disabled.}
\label{tab:denoising_statistics}
        \end{table}

\begin{table}[ht]
           \centering          
\begin{tabularx}{\textwidth}{p{1.5cm}|*{7}{>{\centering\arraybackslash}X}}
        \hline
Failed \acp{TOF} & 1 & 2 & 3 & 1--3 & 2 & 2 & \#8,\#13 \\
scenario & random & random & random & random & neighbors & opposite & position \\
\hline
CAE-32 & 3.27e-3 $\pm$2.51e-3 (-1.88e-3, -1.86e-3) & 3.42e-3 $\pm$2.57e-3 (-1.80e-3, -1.77e-3) & 3.64e-3 $\pm$2.67e-3 (-1.72e-3, -1.69e-3) & 3.47e-3 $\pm$2.61e-3 (-1.81e-3, -1.78e-3) & 3.50e-3 $\pm$2.60e-3 (-1.79e-3, -1.77e-3) & 3.37e-3 $\pm$2.56e-3 (-1.79e-3, -1.77e-3) & 3.40e-3 $\pm$2.56e-3 (-1.79e-3, -1.76e-3) \\
\hline
CAE-64 & 2.48e-3 $\pm$2.11e-3 (-1.09e-3, -1.06e-3) & 2.64e-3 $\pm$2.18e-3 (-1.01e-3, -9.92e-4) & 2.85e-3 $\pm$2.28e-3 (-9.35e-4, -9.11e-4) & 2.68e-3 $\pm$2.21e-3 (-1.02e-3, -9.91e-4) & 2.72e-3 $\pm$2.21e-3 (-1.01e-3, -9.85e-4) & 2.60e-3 $\pm$2.17e-3 (-1.01e-3, -9.90e-4) & 2.60e-3 $\pm$2.17e-3 (-9.89e-4, -9.69e-4) \\
\hline
CAE-128 & 1.93e-3 $\pm$1.60e-3 (-5.33e-4, -5.17e-4) & 2.10e-3 $\pm$1.70e-3 (-4.75e-4, -4.58e-4) & 2.33e-3 $\pm$1.84e-3 (-4.05e-4, -3.85e-4) & 2.14e-3 $\pm$1.75e-3 (-4.71e-4, -4.51e-4) & 2.18e-3 $\pm$1.74e-3 (-4.68e-4, -4.49e-4) & 2.06e-3 $\pm$1.69e-3 (-4.69e-4, -4.52e-4) & 2.06e-3 $\pm$1.69e-3 (-4.47e-4, -4.32e-4) \\
\hline
CAE-256 & 1.54e-3 $\pm$1.24e-3 (-1.40e-4, -1.27e-4) & 1.73e-3 $\pm$1.38e-3 (-1.01e-4, -8.56e-5) & 1.97e-3 $\pm$1.56e-3 (-4.54e-5, -2.58e-5) & 1.77e-3 $\pm$1.44e-3 (-1.01e-4, -8.17e-5) & 1.80e-3 $\pm$1.42e-3 (-8.73e-5, -6.98e-5) & 1.69e-3 $\pm$1.37e-3 (-1.00e-4, -8.50e-5) & 1.72e-3 $\pm$1.36e-3 (-9.79e-5, -8.57e-5) \\
\hline
CAE-512 & \textbf{1.40e-3 $\pm$1.11e-3} & \textbf{1.63e-3 $\pm$1.28e-3} & \textbf{1.93e-3 $\pm$1.49e-3} & \textbf{1.68e-3 $\pm$1.37e-3} & \textbf{1.72e-3 $\pm$1.34e-3} & \textbf{1.59e-3 $\pm$1.27e-3} & \textbf{1.62e-3 $\pm$1.24e-3} \\
\hline
Mean model & 7.77e-2 $\pm$1.23e-2 (-7.64e-2, -7.62e-2) & 7.76e-2 $\pm$1.22e-2 (-7.60e-2, -7.58e-2) & 7.76e-2 $\pm$1.22e-2 (-7.58e-2, -7.56e-2) & 7.77e-2 $\pm$1.22e-2 (-7.62e-2, -7.60e-2) & 7.81e-2 $\pm$1.21e-2 (-7.64e-2, -7.62e-2) & 7.72e-2 $\pm$1.23e-2 (-7.57e-2, -7.55e-2) & 7.77e-2 $\pm$1.22e-2 (-7.62e-2, -7.60e-2) \\
\hline
\end{tabularx}
\caption{Statistics for reconstruction of different failure scenarios across different architectures, where 1 to 3 \ac{TOF} detectors are disabled.}
\label{tab:reconstruction_architectures_statistics}
        \end{table}

\begin{table}[ht]
           \centering    
        \begin{tabularx}{\textwidth}{p{1.5cm}|*{7}{>{\centering\arraybackslash}X}}
        \hline
Failed \acp{TOF} & 1 & 2 & 3 & 1--3 & 2 & 2 & \#8,\#13 \\
scenario & random & random & random & random & neighbors & opposite & position \\
\hline
$\gamma=0.3$ CAE-64 & 2.50e-3 $\pm$2.11e-3 (-3.28e-5, -2.44e-5) & 2.67e-3 $\pm$2.18e-3 (-3.23e-5, -2.09e-5) & 2.89e-3 $\pm$2.28e-3 (-4.30e-5, -2.73e-5) & 2.72e-3 $\pm$2.22e-3 (-4.13e-5, -2.70e-5) & 2.76e-3 $\pm$2.21e-3 (-4.32e-5, -3.00e-5) & 2.63e-3 $\pm$2.17e-3 (-3.77e-5, -2.72e-5) & 2.62e-3 $\pm$2.16e-3 (-2.41e-5, -1.62e-5) \\
\hline
$\gamma=0.7$ CAE-64 & 2.51e-3 $\pm$2.10e-3 (-3.57e-5, -2.72e-5) & 2.67e-3 $\pm$2.18e-3 (-3.15e-5, -2.02e-5) & 2.89e-3 $\pm$2.28e-3 (-3.94e-5, -2.38e-5) & 2.71e-3 $\pm$2.21e-3 (-2.96e-5, -1.54e-5) & 2.75e-3 $\pm$2.21e-3 (-3.16e-5, -1.84e-5) & 2.62e-3 $\pm$2.16e-3 (-3.51e-5, -2.47e-5) & 2.63e-3 $\pm$2.16e-3 (-2.79e-5, -2.00e-5) \\
\hline
$p=0$ CAE-64 & \textbf{2.47e-3 $\pm$2.11e-3} & \textbf{2.64e-3 $\pm$2.18e-3} & \textbf{2.86e-3 $\pm$2.28e-3} & \textbf{2.69e-3 $\pm$2.22e-3} & \textbf{2.72e-3 $\pm$2.21e-3} & \textbf{2.59e-3 $\pm$2.16e-3} & \textbf{2.60e-3 $\pm$2.16e-3} \\
\hline
$p=1$ CAE-64 & 2.52e-3 $\pm$2.11e-3 (-4.53e-5, -3.70e-5) & 2.68e-3 $\pm$2.19e-3 (-5.11e-5, -3.98e-5) & 2.92e-3 $\pm$2.30e-3 (-6.75e-5, -5.16e-5) & 2.74e-3 $\pm$2.23e-3 (-6.26e-5, -4.81e-5) & 2.78e-3 $\pm$2.22e-3 (-6.11e-5, -4.78e-5) & 2.64e-3 $\pm$2.18e-3 (-5.20e-5, -4.15e-5) & 2.65e-3 $\pm$2.18e-3 (-5.00e-5, -4.21e-5) \\
\hline
$p=2$ CAE-64 & 2.56e-3 $\pm$2.15e-3 (-9.06e-5, -8.22e-5) & 2.73e-3 $\pm$2.22e-3 (-9.56e-5, -8.42e-5) & 2.98e-3 $\pm$2.34e-3 (-1.30e-4, -1.14e-4) & 2.79e-3 $\pm$2.27e-3 (-1.09e-4, -9.41e-5) & 2.83e-3 $\pm$2.26e-3 (-1.12e-4, -9.83e-5) & 2.69e-3 $\pm$2.22e-3 (-9.76e-5, -8.70e-5) & 2.68e-3 $\pm$2.21e-3 (-8.64e-5, -7.84e-5) \\
\hline
CCNN & 2.49e-3 $\pm$2.10e-3 (-2.23e-5, -1.39e-5) & 2.65e-3 $\pm$2.17e-3 (-1.52e-5, -3.96e-6) & 2.87e-3 $\pm$2.27e-3 (-1.76e-5, -2.25e-6) & 2.70e-3 $\pm$2.21e-3 (-2.12e-5, -7.04e-6) & 2.73e-3 $\pm$2.21e-3 (-1.73e-5, -4.22e-6) & 2.61e-3 $\pm$2.16e-3 (-2.09e-5, -1.04e-5) & 2.61e-3 $\pm$2.16e-3 (-1.09e-5, -3.02e-6) \\
\hline
\end{tabularx}
\caption{Statistics for reconstruction of different failure scenarios across different parameter settings, where 1 to 3 \ac{TOF} detectors are disabled.}
\label{tab:reconstruction_parameters_statistics}
        \end{table}

\begin{table}[ht]
            \centering    
\begin{tabularx}{\textwidth}{p{1.5cm}|*{7}{>{\centering\arraybackslash}X}}
        \hline
Failed \acp{TOF} & 1 & 2 & 3 & 1--3 & 2 & 2 & \#8,\#13 \\
scenario & random & random & random & random & neighbors & opposite & position \\
\hline
1TOF model & 2.48e-3 $\pm$2.07e-3 (-4.82e-6, 4.28e-6) & 4.00e-3 $\pm$4.05e-3 (-1.40e-3, -1.34e-3) & 6.61e-3 $\pm$6.46e-3 (-3.81e-3, -3.70e-3) & 5.06e-3 $\pm$5.63e-3 (-2.42e-3, -2.33e-3) & 6.72e-3 $\pm$5.67e-3 (-4.04e-3, -3.95e-3) & 2.76e-3 $\pm$2.19e-3 (-1.66e-4, -1.54e-4) & 2.74e-3 $\pm$2.17e-3 (-3.48e-4, -3.37e-4) \\
\hline
2TOF model & 2.64e-3 $\pm$2.22e-3 (-1.73e-4, -1.61e-4) & 2.67e-3 $\pm$2.18e-3 (-4.14e-5, -3.00e-5) & 3.15e-3 $\pm$2.57e-3 (-3.11e-4, -2.85e-4) & 2.90e-3 $\pm$2.48e-3 (-2.28e-4, -2.05e-4) & 2.76e-3 $\pm$2.21e-3 (-4.41e-5, -3.08e-5) & 2.63e-3 $\pm$2.17e-3 (-3.56e-5, -2.51e-5) & 2.63e-3 $\pm$2.16e-3 (-2.33e-4, -2.24e-4) \\
\hline
3TOF model & 3.17e-3 $\pm$5.34e-3 (-7.36e-4, -6.57e-4) & 2.80e-3 $\pm$2.34e-3 (-1.74e-4, -1.58e-4) & 2.86e-3 $\pm$2.28e-3 (-1.48e-5, 1.39e-7) & 2.97e-3 $\pm$3.07e-3 (-3.01e-4, -2.65e-4) & 2.87e-3 $\pm$2.31e-3 (-1.57e-4, -1.42e-4) & 2.78e-3 $\pm$2.32e-3 (-1.94e-4, -1.79e-4) & 2.74e-3 $\pm$2.32e-3 (-3.48e-4, -3.33e-4) \\
\hline
General model & \textbf{2.48e-3 $\pm$2.11e-3} & \textbf{2.64e-3 $\pm$2.18e-3} & \textbf{2.86e-3 $\pm$2.28e-3} & \textbf{2.69e-3 $\pm$2.22e-3} & \textbf{2.72e-3 $\pm$2.21e-3} & \textbf{2.60e-3 $\pm$2.17e-3} & 2.60e-3 $\pm$2.16e-3 (-2.06e-4, -1.97e-4) \\
\hline
Spec model & 9.99e-3 $\pm$7.11e-3 (-7.58e-3, -7.45e-3) & 1.36e-2 $\pm$8.64e-3 (-1.10e-2, -1.09e-2) & 1.68e-2 $\pm$9.50e-3 (-1.40e-2, -1.38e-2) & 1.34e-2 $\pm$9.22e-3 (-1.08e-2, -1.06e-2) & 1.35e-2 $\pm$8.77e-3 (-1.09e-2, -1.07e-2) & 1.34e-2 $\pm$9.13e-3 (-1.09e-2, -1.08e-2) & \textbf{2.40e-3 $\pm$2.00e-3} \\
\hline
Mean model & 7.76e-2 $\pm$1.23e-2 (-7.53e-2, -7.51e-2) & 7.75e-2 $\pm$1.22e-2 (-7.50e-2, -7.48e-2) & 7.76e-2 $\pm$1.22e-2 (-7.48e-2, -7.46e-2) & 7.77e-2 $\pm$1.22e-2 (-7.51e-2, -7.49e-2) & 7.81e-2 $\pm$1.21e-2 (-7.55e-2, -7.53e-2) & 7.73e-2 $\pm$1.23e-2 (-7.48e-2, -7.46e-2) & 7.78e-2 $\pm$1.21e-2 (-7.54e-2, -7.53e-2) \\
\hline
\end{tabularx}

\caption{Statistics for reconstruction of different failure scenarios, where 1 to 3 \ac{TOF} detectors are disabled.}
\label{tab:failure_statistics}
        \end{table}

\begin{table}[ht]
            \centering    
        \begin{tabularx}{\textwidth}{p{1.5cm}|*{4}{>{\centering\arraybackslash}X}}
        \hline
Failed \acp{TOF} & 4 & 5 & 1--4 & 1--5 \\
scenario & random & random & random & random \\
\hline
CAE-64 & 3.17e-3 $\pm$2.46e-3 (-6.40e-5, -4.28e-5) & 3.63e-3 $\pm$2.79e-3 (-2.03e-4, -1.72e-4) & \textbf{2.85e-3 $\pm$2.33e-3} & 3.06e-3 $\pm$2.55e-3 (-6.98e-5, -4.31e-5) \\
\hline
1-4TOF & 3.14e-3 $\pm$2.42e-3 (-3.12e-5, -1.16e-5) & 3.50e-3 $\pm$2.64e-3 (-7.46e-5, -4.79e-5) & 2.86e-3 $\pm$2.31e-3 (-1.86e-5, 7.59e-7) & 3.03e-3 $\pm$2.45e-3 (-3.25e-5, -9.16e-6) \\
\hline
1-5TOF & \textbf{3.12e-3 $\pm$2.41e-3} & \textbf{3.44e-3 $\pm$2.59e-3} & 2.86e-3 $\pm$2.31e-3 (-2.09e-5, -1.87e-6) & \textbf{3.01e-3 $\pm$2.42e-3} \\
\hline
Mean model & 7.77e-2 $\pm$1.22e-2 (-7.47e-2, -7.45e-2) & 7.82e-2 $\pm$1.22e-2 (-7.48e-2, -7.46e-2) & 7.79e-2 $\pm$1.22e-2 (-7.52e-2, -7.50e-2) & 7.81e-2 $\pm$1.21e-2 (-7.52e-2, -7.50e-2) \\
\hline
\end{tabularx}

\caption{Statistics for reconstruction of different failure scenarios, where 4 or 5 \ac{TOF} detectors are disabled.}
\label{tab:failure_statistics_4_5}
        \end{table}
\clearpage
\section{Checking for Saturation of \texorpdfstring{\ac{TOF}}{TOF} detectors}
\label{sec:saturation}
\begin{figure}
    \centering
    \includegraphics[width=0.8\linewidth]{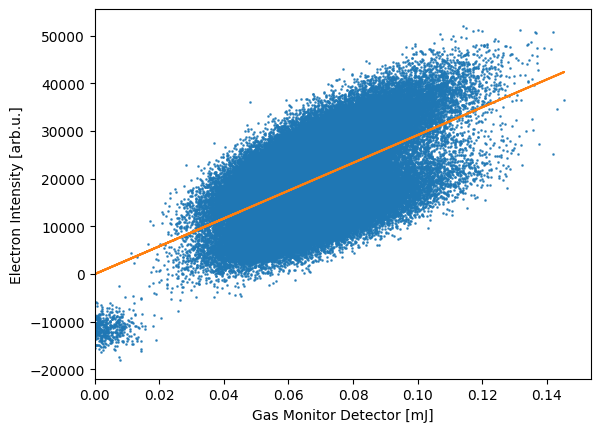}
    \caption{Electron intensities vs. gas monitor detector values. The blue dots represent the measured data points, while the orange line is the linear approximation of these points. The plot is baseline corrected.}
    \label{fig:saturation}
\end{figure}

In this section, we compare the electron intensities measured by all \ac{TOF} detectors with the outputs from the gas monitor detector. The gas monitor detector measures the calibrated ionization count, reflecting the overall ionization produced by the photons. As shown in \cref{fig:saturation}, the blue data points in the plot remain linearly increasing in the higher regions of the gas monitor detector, closely following the linear approximation (orange line). This behavior suggests that the \ac{TOF} detectors do not show saturation. For the calculation of the linear approximation, data points near $(0,0)$ were excluded, as these correspond to shots recorded without lasing. For this scenario, both the gas monitor detector and the \ac{TOF} detectors are uncalibrated and return values close to $(0,0)$, making them unsuitable for reliable analysis. We corrected the baseline of the plot by shifting the linear fit to pass through the origin. Consequently, the data points were also adjusted to reflect this baseline correction.

\clearpage

\section{\ac{TOF} Detector Disablement Algorithm}
\begin{algorithm}[H]
\caption{Algorithm for selecting random \ac{TOF} detectors to disable, ensures to show samples with neighboring and opposite failed \ac{TOF} detectors more often, as these constellations are harder to reconstruct.}\label{alg:random_tofs}
\begin{algorithmic}[1]
	\Procedure{RandomTOFs}{$n_{max}, \xi$}\Comment{$n_{max}$ is the maximum amount of TOFs to disable, $\xi$ is the neighbor or opposite probability}
	\State $l_{TOF} \gets \pi \left( \left[ 1, \dots, n_{TOF} \right] \right) $ 
	\Comment{$n_{TOF}$ is the total amount of TOFs, $\pi$ is a random permutation}
 \State $l_{disabled} \gets \left[ l_{TOF}[0] \right] $
	\State $n_{disabled} \sim \mathcal{U}\left(  0, n_{max}  \right) $
	\For{$i \in \left\{ 0, \dots, n_{disabled} \right\}$ }
	    \State $r \sim \mathcal{U}\left(0, 1\right)$
	\If{$r < \xi$}
 
	\If{$r < \frac{\xi}{2} $}\Comment{Look for neighbor}
 \If{for one element in $l_{disabled}$ the left or right neighbor $\eta$ is in $l_{TOF}$}
 \State AddItem($l_{disabled}$, $\eta$)
 \State RemoveItem($l_{TOF}$, $\eta$)
 \Else     \Comment{If no neighbor, add a random element}
    \State AddItem($l_{disabled}$, $l_{TOF}[0]$)
 \EndIf
 \Else \Comment{Look for opposite}
  \If{for one element in $l_{disabled}$ the opposite $\omega$ is in $l_{TOF}$}
 \State AddItem($l_{disabled}$, $\omega$)
 \State RemoveItem($l_{TOF}$, $\omega$)
 \Else  \Comment{If no opposite, add a random element}
        \State AddItem($l_{disabled}$, $l_{TOF}[0]$)
       
 \EndIf
	\EndIf
	\EndIf
	\EndFor
\State \textbf{return} $l_{disabled}$
\EndProcedure
\end{algorithmic}
\end{algorithm}

\section{Impact of Regularization}
\label{ssec:regularization_impact}

\begin{center}
\begin{figure*}
    \centering
\includegraphics[width=\textwidth]{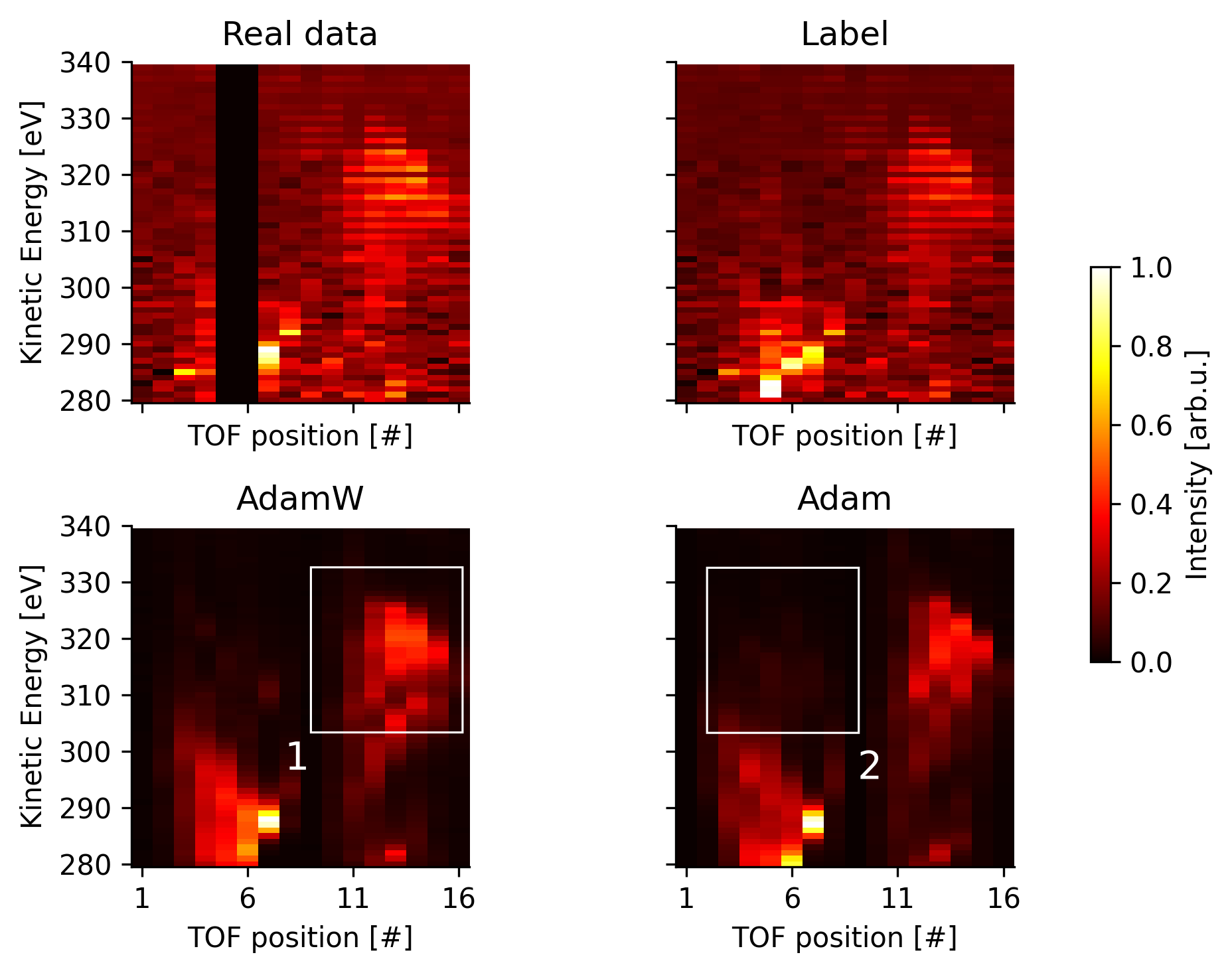}
\caption{
Visual comparison of reconstruction performance between a network trained with AdamW (with regularization) and  Adam (without regularization). }
\label{fig:adam_w_comparison}
\end{figure*}
\end{center}

As shown in \cref{fig:adam_w_comparison}, the AdamW-trained network reconstructs the broader intensity area more accurately (denoted with 1). In the region with minimal intensity in the label, the Adam-trained network reconstructs more incorrect intensity than the AdamW-trained network (2).

\section{Phase-Separated Evaluation}
\label{ssec:phase_evaluation}

\begin{center}
\begin{figure*}
    \centering
\includegraphics[width=\textwidth]{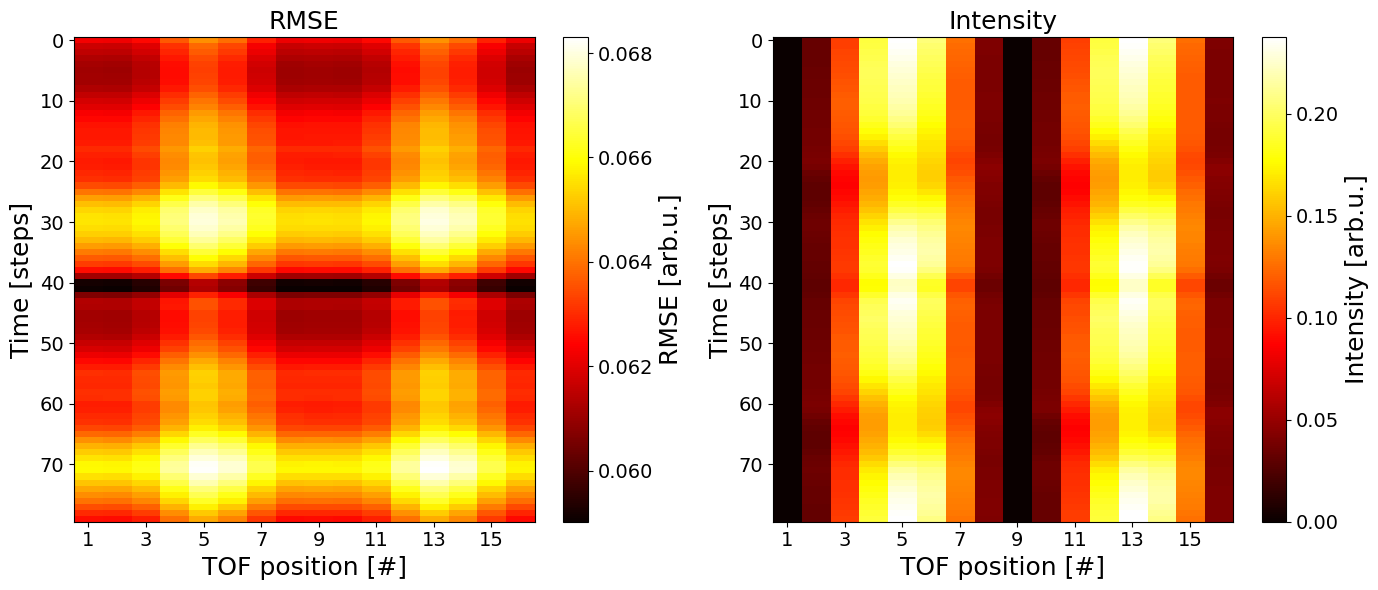}
\caption{
Phase-resolved analysis of reconstruction quality. The left plot depicts the \ac{RMSE} of the reconstruction for each phase step when a single \ac{TOF} detector is disabled. The right plot shows the square root of the average intensities of 10,000 detector images across all phases and \ac{TOF} positions.}
\label{fig:phase_evaluation}
\end{figure*}
\end{center}

In this section, we analyze the reconstruction quality across different phases. To achieve this, we generate 10\,000 spectrograms spanning $80$ phase steps, each with a single intensity peak at the corresponding phase and a random photon energy. Noise and other parameters are set accordingly to \cref{sec:tof_rec_method}. These spectrograms are then transformed into detector images for evaluation.

Using the General model, we assess the detector images by systematically deactivating one \ac{TOF} detector at a time and calculating the \ac{RMSE} of the resulting reconstructions. The results are presented in the left plot of \cref{fig:phase_evaluation}, which shows that certain phases consistently exhibit lower reconstruction errors. For instance, phases around step 40 demonstrate higher reconstruction accuracy. This improvement arises because the maxima of the corresponding partial waves align with the maxima of the angular distribution, as illustrated in \cref{fig:1_tof_failed}.

Contrary, specific combinations of phases and \ac{TOF} detectors result in increased reconstruction errors. Especially phase steps 30 or 70, combined with a failure of \ac{TOF} positions \#5 or \#13. These combinations lead to a significant loss of information regarding the relevant partial waves, thus degrading the reconstruction accuracy. For reference, we plotted the square root of the average intensities of all 10\,000 detector images across all phases and \ac{TOF} positions on the right side of \cref{fig:phase_evaluation}.

\section{Wiener Filter}
\label{sec:wiener}
The Wiener Filter in our used implementation by \citeauthor{Virtanen2020} works by calculating the following steps \cite{Virtanen2020, Lim1990}:

\begin{enumerate}
\item \textbf{Local Mean} $\mu$

\[
\mu(x) = \frac{1}{|N|} \sum_{i \in N} I(x + i)
\]

where $I$ is the input detector image, $N$ is in our case a $(3,3)$ part from the entire detector image, referred to as window, $|N|$ is the total number of elements in the window $N$, and $x$ represents the current pixel or element location.

\item \textbf{Local Variance} $\sigma^2$

\[
\sigma^2(x) = \frac{1}{|N|} \sum_{i \in N} I(x + i)^2 - \mu(x)^2
\]

\item \textbf{Noise Power} $P_{\text{noise}}$

The noise power is estimated as the mean of the local variances:

\[
P_{\text{noise}} = \frac{1}{|X|} \sum_{x \in X} \sigma^2(x)
\]

where $|X|$ is the total number of elements in $I$.

\item \textbf{Filtered Output} $O(x)$

Compute the Wiener filtered output:

\[
O(x) =
\begin{cases}
\mu(x), & \text{if } \sigma^2(x) < P_{\text{noise}}, \\
\mu(x) + \left(1 - \frac{P_{\text{noise}}}{\sigma^2(x)}\right) \cdot \left(I(x) - \mu(x)\right), & \text{otherwise.}
\end{cases}
\]
\end{enumerate}

This process is applied to every data point $x$ in the input array $I$.
\section{Data Availability}
All dataset generation and program scripts to this article can be found at a repository hosted at Github: \url{https://github.com/hz-b/tof-reconstruction}.

\bibliography{main}

\end{document}